\documentclass{article}
\usepackage{amsmath,amsthm,amssymb}
\usepackage[square,numbers]{natbib}
\usepackage[latin1]{inputenc}
\usepackage[T1]{fontenc}
\usepackage{url}
\usepackage{color}
\usepackage{epsfig}
\usepackage[normalem]{ulem}
\usepackage{listings}
\usepackage{tikz}
\usepackage{float}
\usepackage{lscape}
\usepackage{hyperref}
\usetikzlibrary{arrows,decorations.pathmorphing,backgrounds,fit,positioning,shapes.symbols,chains}
\usetikzlibrary{decorations.pathmorphing}
\usetikzlibrary{arrows,shapes,trees}

\definecolor{dgreen}{rgb}{0,.6,0}
\newtheorem*{principle}{Principle}
\newtheorem{definition}{Definition}

\graphicspath{{figures/}}

\title{Formal security analysis of registration protocols for interactive systems: a methodology and a case of study}
\author{Jesus Diaz, David Arroyo, Francisco B. Rodriguez}

\begin{document}
\maketitle

  \begin{abstract}
    In this work we present and formally analyze CHAT-SRP (\textbf{CHA}os based \textbf{T}ickets-\textbf{S}ecure 
    \textbf{R}egistration \textbf{P}rotocol), a protocol to provide interactive 
    and collaborative platforms with a cryptographically robust solution to classical security 
    issues. Namely, we focus on the secrecy and authenticity properties
    while keeping a high usability. In this sense, users are forced to blindly
    trust the system administrators and developers. Moreover, as far as we know,
    the use of formal methodologies for the verification of 
    security properties of communication protocols isn't
    yet a common practice. We propose here a methodology to fill this gap, i.e., 
    to analyse both the security of the proposed protocol and the pertinence of the underlying 
    premises. In this concern, we propose the definition and formal evaluation of a protocol 
    for the distribution of digital identities. Once distributed, these identities 
    can be used to verify integrity and source of information. We base our security analysis 
    on tools for automatic verification of security protocols widely accepted by the scientific 
    community, and on the principles they are based upon. In addition, it is assumed 
    perfect cryptographic primitives in order to focus the analysis on the exchange of protocol
    messages. The main property of our protocol is 
    the incorporation of \emph{tickets}, created using digests of chaos based \emph{nonces} 
    (numbers used only once) and users' personal data. Combined with a multichannel authentication
    scheme with some previous knowledge, these tickets provide security during 
    the whole protocol by univocally linking each 
    registering user with a single request. This way,
    we prevent impersonation and Man In The Middle attacks, which are the main security problems 
    in registration protocols for interactive platforms. As
    a proof of concept, we also present the results obtained after testing this protocol with
    real users, at our university, in order to measure the usability of the registration
    system.
  \end{abstract}

\section{Introduction}
\label{sec:introduction}

% Motivation
Modularity is one of the most relevant aspect of modern engineering 
practices. Regarding cryptographic applications, this property resorts to 
distinguishing the formal aspect of cryptographic protocols from the inner 
details of the underlying algorithms (see \citep{ag00,ghrs05}). In this vein, 
we can design strong cryptographic algorithms for confidentiality, integrity and authentication, 
but applying them incorrectly would lead to security flaws. 
Therefore, the study of the security of a  cryptographic protocol demands to 
examine the security of cryptographic primitives from a computational point of
view, but also  to evaluate the goodness of the integration of those primitives.
In this work we propose a registration protocol for interactive platforms. 

The proper design and evaluation of cryptograpic protocols is critical 
when personal information is exchanged. This is the case of interactive and 
collaborative platforms, which are of great importance in the current 
state of communications, specially after the irruption of web 2.0  technologies. 
This type of applications are used to share and exchange information, even 
critical personal data \citep{wg08}. Nonetheless, users' trust is generally implicitly assumed
and there is no explicit application of procedures to secure users' registration 
and exchange of information. In the context of interactive applications, and
always from a general point of view, the main properties of the underlying
security system rely completely on the correct implementation and subsequent 
management of the system. Privacy, secrecy and authenticity are assumed when 
users get into the system, i.e., by adhering straightway to solutions provided by
administrators and developers. In this sense, the only way to incorporate the basics 
of information security is through properly using standard tools proposed, 
evaluated and validated by the cryptography and information security community. 

% Our contribution
%% What is it and what does it provide?

On the regard of applying standard and validated tecnologies, we have previously
introduced a registration protocol to enhance interactive platforms' security \citep{dar11}. 
This registration protocol links each user to a digital identity, giving users 
access to the cryptographic tools sustaining confidentiality and authenticity, i.e., 
(client-side) encryption and digital signatures. The service provided by our 
registration protocol could seem similar to the protection given by SSL-tunneled 
communications. Nevertheless, it rather complements SSL
instead of overlapping its functionality: SSL provides secrecy and server side 
authentication, whilst client-side authentication cannot be provided by SSL as 
long as the client does not have a digital identity, which he would have
once incorporated our protocol into the system. Moreover, thanks to
the distribution of digital identities, and the cryptographic functionality they
make available, their information may be also protected in the servers, 
since it can be encrypted or signed at client side. Therefore
the sensitive information will be 
protected not only during the communications. And even more, this way, users  
will have a greater (and justified) sensation of security, which contributes to preserve a very 
important property (when it is correctly grounded) of information 
security systems:  users' confidence in the system.

In this work we further refine the previous protocol, and formally analyse its
security properties. Our protocol is based on \textbf{E}mail \textbf{B}ased 
\textbf{I}dentification and \textbf{A}uthentication (EBIA), which
links each new user to an email account. This kind of registration protocols for
interactive platforms is widely extended in the internet. Nevertheless, 
there are two main security problems with these protocols: 
impersonation and Man In The Middle (MITM) attacks. We tackle these problems here,
proposing methods for circumventing them.

%% Methodology

In any case, for guaranteeing that the ultimately distributed digital 
identities will successfully provide all their functionality, we need to ensure 
that we leave no unnoticed ``security holes'' in the 
registration protocol. To the best of our knowledge, and from a general point of view, implementation 
and design of registration protocols are not evaluated according to a formal methodology
and using specialized tools intended for that end, like \cite{cryptyc,blanchet10}.
In our work, we address this problem incorporating a phase-divided methodology, and
using formal tools (in our case, ProVerif \cite{blanchet10}) for the verification of the required 
security properties. It is worth to emphasize that we restrain ourselves to
analyzing the security of the protocol itself, i.e., of the exchange of messages, while
assuming a perfect cryptography model for all the cryptographic primitives used.
Additionally, the final step of our work is on testing the usability 
of the protocol, in practice. In order to test it in a real scenario, we have incorporated 
the protocol in a Moodle platform. Moodle is probably the most extended e-learning 
platform worldwide nowadays \citep{app10}. It is also a perfect example
of interactive and collaborative platform managing lots of sensitive data. 
But even more, EBIA is one of the most used authentication modules when working with Moodle.
Therefore, it is a suitable context to test our protocol in.

% Article structure
The rest of the paper is organized as follows. In Section \ref{sec:securityreg} we
make an outline of the basic registration protocols for interactive platforms and
expose their main security problems. In Section \ref{sec:methodology}
we explain the methodology followed during our work. Afterwards, in Section \ref{sec:protocol} 
we introduce our new protocol, covering the first four steps of the procedure
proposed in Section \ref{sec:methodology} and justifying the decissions
that have led us to its final form. Section \ref{sec:analysis} is dedicated to the 
last step of our methodology, centered on the formal verification of the protocol. In Section 
\ref{sec:analysis} we also treat the usability properties of CHAT-SRP, showing the results 
obtained during tests performed with real users. At last, we conclude in Section 
\ref{sec:conclusion}, with an overview of our work and some discussions on 
future work.

\section{A brief security analysis of interactive registration protocols}
\label{sec:securityreg}

\begin{figure}[ht!]
\begin{center}
\includegraphics[width=0.80\textwidth]{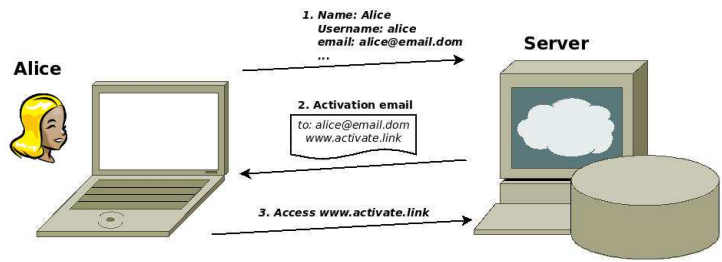}
\caption{Behaviour of EBIA systems\label{fig:ebia}}
\end{center}
(1) The user sends the request with his data. 
(2) The server validates the data and sends an activation link via email. (3) The
user accesses the activation link.
\end{figure}

Our registration protocol takes as starting point the \emph{EBIA} approach, %% How it works: basis and main properties
reinforcing it with robust cryptographic functionality to guarantee secrecy and authenticity.
\emph{EBIA} is the most widely used registration and authentication system in interactive
and collaborative systems (see \citep{garfinkel03}). The reason of taking it as
starting point is its high expansion and usability, as every user of 
the internet is accustomed to its principles. Basically, for a given email 
address, e.g. \emph{alice@email.dom}, EBIA says that if somebody can read an email 
sent to \emph{alice@email.dom}, then she/he is the legitimate owner of that email 
account. As a result, EBIA \emph{assigns} her/his virtual identity to that email 
address. During the registration process, the user has to access an activation link 
sent in an email in order to activate the account (this process is schematized in 
Fig. \ref{fig:ebia}).

\begin{figure}[ht!]
\begin{center}
\includegraphics[width=0.80\textwidth]{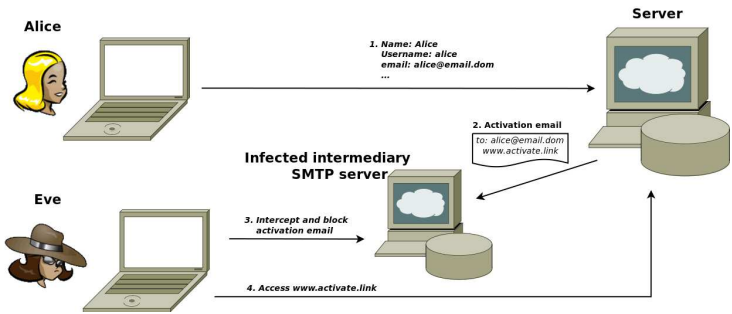}
\caption{Attack to EBIA system. 
  (1) Alice requests registration to the Web Server; 
  (2) The server sends the activation email to Alice, containing the activation
      link, which passes through an infected intermediary server under Eve's control.
  (3) Eve intercepts and blocks the activation email, and activates the
      account in Alice's behalf. Alice will probably just think that an error 
      occurred.\label{fig:ebiaattack}}
\end{center}
\end{figure}

Such a registration process presents two main security flaws. First, the activation email 
is sent unencrypted, which makes possible to mount a MITM attack
(see \citep[Chapter 2]{anderson08}). Let us 
consider the illustrative example shown in Fig. \ref{fig:ebiaattack}. If the
attacker Eve wants to impersonate Alice in a web site using EBIA, she can proceed 
as follows: first, Eve waits until Alice sends a registration request to the Web 
Server (WS); then, the WS will send an activation email to Alice's email account; 
after gaining control of an intermediary server, Eve intercepts the unencrypted
email, gets the activation link, and blocks the email impeding to reach Alice's email account; at last, Eve just 
needs to access the link in order to successfully complete the impersonation attack.
As Alice did not receive an email, she will probably just think that an error occurred. As we 
discuss later, we use registration tickets for ensuring that whoever starts a registration 
request is the one obtaining the corresponding digital identity. 

Now, let us assume that the previous problem is indeed solved. Then, the second problem 
 is that
if the attacker knows all the required data for a user to be registered, he can 
successfully impersonate him from the beginning. To the best of our knowledge, 
the easiest (and maybe only) solution to this problem is to link identities
distribution with some previous information known by the registrar about the users. In this respect, 
and before the registration request, the registrar must possess some information 
concerning the user. This imposes very strong limitations to the contexts in 
which the resulting registration protocol will be suitable for. Nevertheless,
there are many situations in which this is not an unreasonable requirement. In 
our case, we will require the registrar to know the mobile phone number of the 
users. With it, we will be able to perform a multichannel protocol with an 
authenticated channel (the mobile phone), which guarantees that the user 
completing registration is who claims to be.

\section{Methodology for design and verification of secure protocols}
\label{sec:methodology}

\begin{figure}[ht!]
  \begin{center}
    \includegraphics[width=0.95\textwidth]{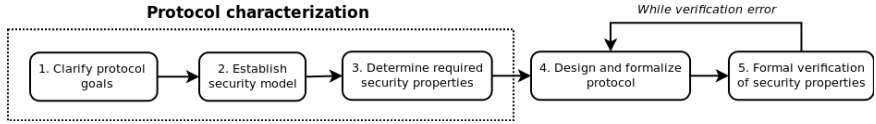}
    \caption{Proposed procedure for the creation and design of secure
      protocols: (1) Make clear the protocol goals; (2) Establish the
      security model and assumptions; (3) Determine the required 
      security properties for achieving the goals; (4) Design and
      formalize the protocol; (5) Carry out a formal verification of
      the required security properties.\label{fig:procedure}}
  \end{center}
\end{figure}

When considering implementation and design of registration protocols, the most 
common practice is adhering to a security model, design the protocol, and informally
claim that the protocol is secure according to the assumed security model. In other 
words, security is assumed instead of being formally tested. The lack of a formal
methodology to evaluate security leads to unnoticed errors, which may later cause 
severe damage to the protocols or even make them useless. Consequently, it is highly
convenient to adopt methodologies based on the formal analysis of protocols, as 
the one applied in \citep{wkv11} for verification of an electronic voting system. 
In this vein, we propose the procedure schematized in Fig. \ref{fig:procedure}, 
which consists of five stages. The first stage  define the goals of our protocol, 
enabling the identification of critical aspects and the complexity of the associated 
problem. After this initial analysis, we have enough information to concrete an
abstract model in coherence with the practical scenario where the protocol will 
run (second stage of the methodology). According to the security abstract model, 
in the third stage we formally define the security properties and discuss their 
feasibility. Sometimes, this ``preliminary'' analysis is omitted because we
think that we perfectly know our context, and it is not worth the effort. 
However, as it is said in \cite{kbg11}, it is desirable and beneficial
to follow a methodological approach for establishing the desired security goals and 
requirements. Finally, the protocol is designed and formalized in order to be 
analyzed with automatic and proved tools (stages four and five).

It can be seen that the steps of our methodology follow a natural order. As far 
as we know, steps one and, sometimes, three are usually ignored, which can involve contradictions 
and/or disregarding of the relevance of some property. The second step, although 
it may be commonly applied in any explicit way, helps modelling the scenario and 
determining the security properties. These three first steps can be
summarized in a more general phase: the protocol (and environment) characterization. Nevertheless
it is worth separating them, since although related, they are intended for 
different aims. The fourth step could be introduced within 
the fifth one (since formal verification of the protocol demands its previous 
formalization), but it is worth to be considered separately, since improves the 
designer's knowledge of the protocol. Note that we have included a
loop back from the fifth to the fourth step, in case some security property is
not held (we assume that if a security property is not possible in a given 
environment, it will be detected in the first three steps).

As we will see in the subsequent sections, we have followed this methodology for the
design and evaluation of our protocol.

\section{The proposed protocol: CHAT-SRP}
\label{sec:protocol}

In this section we present CHAT-SRP, our proposed registration protocol. We
introduce here the first four steps of the proposed methodology, reasoning
about why the main modifications over EBIA are necessary. We leave the 
formal security analysis for the following sections, where we will
show that the impersonation and MITM attacks explained before are
avoided.

\subsection{Protocol characterization: goals, security model and security requirements}
\label{ssec:goals}

CHAT-SRP is a registration protocol mainly intended for interactive and collaborative 
platforms. Therefore, its goal is to provide new users with digital identities. These 
digital identities are intented to be used for cryptographic purposes, like encryption 
and digital signatures. Obviously, the user will take active part in the protocol.
Also, several servers will interact between them during registration.

Regarding the assumed security model, we have based our analysis upon the
Dolev-Yao model \citep{dy83}. In it, the cryptographic 
primitives used are supposed to be perfectly secure, i.e., the attacker is 
not able to decipher the encrypted messages unless he has the corresponding 
decryption key,  
the random number generators create unpredictable random numbers, etc. Nevertheless, the attacker is assumed to be active, 
which means that he can intercept, resend, and insert messages. We take into
account both external and internal attackers (see \citep{diaz05} for
definitions concerning the \emph{location} of the attacker). In order to model internal
attackers, we allow the establishment of SSL sessions with the trusted third parties
of the system. This way, an attacker has the capability to act from the inside.

A registration protocol is really an authentication protocol executed for a first time. 
As it is stated in
\citep{wl93}, authentication protocols require two main properties: authentication
and key distribution. The authentication property refers to being certain of the
identity of the users, i.e., of the authenticity of \emph{all} the data they provide,
including their identities. The key distribution also deals with secrecy, assuring that the 
new users will be able to communicate with the rest of the principals preserving the
confidenciality of the information transmitted through the network. Therefore, we will 
incorportate authenticity and secrecy into CHAT-SRP as a commitment.

\subsection{Description of the protocol}
\label{ssec:informal}

Now we have covered the first three steps of the procedure (see Fig. 
\ref{fig:procedure}). But before undertaking the fourth one, 
we explain in more depth the protocol internals. Once we have gained an in-depth 
knowledge of it, we will proceed with the formalization.

The principals involved in the protocol are four:
\begin{itemize}
  \item The User, which is the one starting the protocol by asking for a new digital
    identity linked to her/his email.
  \item The Web Server (WS), which is the entity attending registration requests and 
    acting as intermediary
    between the User and the Registration Authority. In addition, it generates the 
    activation email, and performs some easy checks. \emph{This is a trusted server!}
  \item The Registration Authority (RA), which is on charge of creating the tickets,
    i.e., of linking each user with a single ticket. \emph{This is a trusted server!}
  \item The Certification Authority (CA), that creates the final digital identities upon
    requests of the RA. \emph{This is a trusted server!}
\end{itemize}

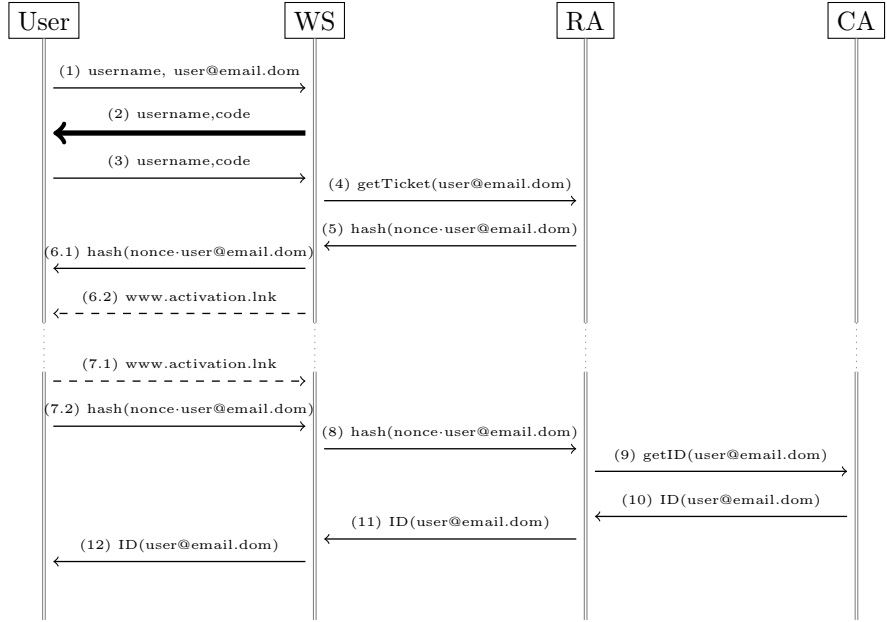
\begin{figure}[!h]
  \begin{center}
    % Sequence diagram for moodle_secure_verif.tex

\begin{tikzpicture}[scale=1.2]
  
  % Nodes
  \node (user) at (1,7.75) [draw] {User};
  \node (user1) at (1,7.00) [draw=none] {};
  \node (user2) at (1,6.50) [draw=none] {};
  \node (user3) at (1,6.00) [draw=none] {};
  \node (user4) at (1,5.75) [draw=none] {};
  \node (user5) at (1,5.25) [draw=none] {};
  \node (user6) at (1,5) [draw=none] {};
  \node (user7) at (1,4.5) [draw=none] {};
  \node (user8) at (1,3.75) [draw=none] {};
  \node (user9) at (1,3.25) [draw=none] {};
  \node (user10) at (1,3) [draw=none] {};
  \node (user11) at (1,2.75) [draw=none] {};
  \node (user12) at (1,2.25) [draw=none] {};
  \node (user13) at (1,2) [draw=none] {};
  \node (user14) at (1,1.75) [draw=none] {};
  \node (enduser) at (1,1) {};
  
  \node (ws) at (4,7.75) [draw] {WS};
  \node (ws1) at (4,7.00) [draw=none] {};
  \node (ws2) at (4,6.50) [draw=none] {};
  \node (ws3) at (4,6.00) [draw=none] {};
  \node (ws4) at (4,5.75) [draw=none] {};
  \node (ws5) at (4,5.25) [draw=none] {};
  \node (ws6) at (4,5) [draw=none] {};
  \node (ws7) at (4,4.5) [draw=none] {};
  \node (ws8) at (4,3.75) [draw=none] {};
  \node (ws9) at (4,3.25) [draw=none] {};
  \node (ws10) at (4,3) [draw=none] {};
  \node (ws11) at (4,2.75) [draw=none] {};
  \node (ws12) at (4,2.25) [draw=none] {};
  \node (ws13) at (4,2) [draw=none] {};
  \node (ws14) at (4,1.75) [draw=none] {};
  \node (endws) at (4,1) {};

  \node (ra) at (7,7.75) [draw] {RA};
  \node (ra1) at (7,7.00) [draw=none] {};
  \node (ra2) at (7,6.50) [draw=none] {};
  \node (ra3) at (7,6.00) [draw=none] {};
  \node (ra4) at (7,5.75) [draw=none] {};
  \node (ra5) at (7,5.25) [draw=none] {};
  \node (ra6) at (7,4.75) [draw=none] {};
  \node (ra7) at (7,4.5) [draw=none] {};
  \node (ra8) at (7,3.75) [draw=none] {};
  \node (ra9) at (7,3.25) [draw=none] {};
  \node (ra10) at (7,3) [draw=none] {};
  \node (ra11) at (7,2.75) [draw=none] {};
  \node (ra12) at (7,2.25) [draw=none] {};
  \node (ra13) at (7,2) [draw=none] {};
  \node (ra14) at (7,1.75) [draw=none] {};
  \node (endra) at (7,1) {};

  \node (ca) at (10,7.75) [draw] {CA};
  \node (ca1) at (10,7.00) [draw=none] {};
  \node (ca2) at (10,6.50) [draw=none] {};
  \node (ca3) at (10,6.00) [draw=none] {};
  \node (ca4) at (10,5.75) [draw=none] {};
  \node (ca5) at (10,5.25) [draw=none] {};
  \node (ca6) at (10,4.75) [draw=none] {};
  \node (ca7) at (10,4.5) [draw=none] {};
  \node (ca8) at (10,3.75) [draw=none] {};
  \node (ca9) at (10,3.25) [draw=none] {};
  \node (ca10) at (10,3) [draw=none] {};
  \node (ca11) at (10,2.75) [draw=none] {};
  \node (ca12) at (10,2.25) [draw=none] {};
  \node (ca13) at (10,2) [draw=none] {};
  \node (ca14) at (10,1.75) [draw=none] {};
  \node (endca) at (10,1) {};

  % Life lines
  \draw[-] (user.south) -- (user7.south) [double,color=gray];
  \draw[-] (user7.south) -- (user8.north) [dotted,color=gray];
  \draw[-] (user8.north) -- (enduser.north) [double,color=gray];
  \draw[-] (ws.south) -- (ws7.south) [double,color=gray];
  \draw[-] (ws7.south) -- (ws8.north) [dotted,color=gray];
  \draw[-] (ws8.north) -- (endws.north) [double,color=gray];
  \draw[-] (ra.south) -- (ra7.south) [double,color=gray];
  \draw[-] (ra7.south) -- (ra8.north) [dotted,color=gray];
  \draw[-] (ra8.north) -- (endra.north) [double,color=gray];
  \draw[-] (ca.south) -- (ca7.south) [double,color=gray];
  \draw[-] (ca7.south) -- (ca8.north) [dotted,color=gray];
  \draw[-] (ca8.north) -- (endca.north) [double,color=gray];

  % Messages
  \draw[->] (user1) -- (ws1) [line width=0.5pt] node [midway,above,draw=none]{\tiny{(1) username, user@email.dom}};
  \draw[->] (ws2) -- (user2) [line width=2pt] node [midway,above,draw=none]{\tiny{(2) username,code}};
  \draw[->] (user3) -- (ws3) [line width=0.5pt] node [midway,above,draw=none]{\tiny{(3) username,code}};
  \draw[->] (ws4) -- (ra4) [line width=0.5pt] node [midway,above,draw=none]{\tiny{(4) getTicket(user@email.dom)}};
  \draw[->] (ra5) -- (ws5) [line width=0.5pt] node [midway,above,draw=none]{\tiny{(5) hash(nonce$\cdot$user@email.dom)}};
  \draw[->] (ws6) -- (user6) [line width=0.5pt] node [midway,above,draw=none]{\tiny{(6.1) hash(nonce$\cdot$user@email.dom)}};
  \draw[->] (ws7) -- (user7) [line width=0.5pt,dashed] node [midway,above,draw=none]{\tiny{(6.2) www.activation.lnk}};
  \draw[->] (user8) -- (ws8) [line width=0.5pt,dashed] node [midway,above,draw=none]{\tiny{(7.1) www.activation.lnk}};
  \draw[->] (user9) -- (ws9) [line width=0.5pt] node [midway,above,draw=none]{\tiny{(7.2) hash(nonce$\cdot$user@email.dom)}};
  \draw[->] (ws10) -- (ra10) [line width=0.5pt] node [midway,above,draw=none]{\tiny{(8) hash(nonce$\cdot$user@email.dom)}};
  \draw[->] (ra11) -- (ca11) [line width=0.5pt] node [midway,above,draw=none]{\tiny{(9) getID(user@email.dom)}};
  \draw[->] (ca12) -- (ra12) [line width=0.5pt] node [midway,above,draw=none]{\tiny{(10) ID(user@email.dom)}};
  \draw[->] (ra13) -- (ws13) [line width=0.5pt] node [midway,above,draw=none]{\tiny{(11) ID(user@email.dom)}};
  \draw[->] (ws14) -- (user14) [line width=0.5pt] node [midway,above,draw=none]{\tiny{(12) ID(user@email.dom)}};

  % Balloon
  %  \node (balloon) at (8.5,7) {};

  % Balloon edge
  % \path (9.north) edge [] node {} (balloon);
  
\end{tikzpicture}
    \caption{Sequence diagram of the messages exchanged during the protocol. The dashed
      lines represent unprotected communications; the thin continuous ones represent SSL
      protected communications, and the thick continuous line represents the message sent
      using the extra authenticated channel (SMS):
      (1) The user requests registration, providing his personal identification data;
      (2) The WS sends a code to the mobile phone number associated to the requesting user;
      (3) The user provides the recevied code to the WS;
      (4) The WS forwards the request to the RA;
      (5) The RA generates a ticket for the user, and sends it to the WS;
      (6.1) The WS forwards the ticket to the user via SSL;
      (6.2) The WS generates an activation link and sends it via SMTP to the user;
      (7.1) The user accesses the activation link;
      (7.2) The users sends the ticket to the WS;
      (8) After verifying the activation link, the WS forwards the ticket to the RA;
      (9) After verifying the ticket and deleting it, the RA requests the CA to issue
          a  new digital identity;
      (10) The CA generates the user's digital identity and sends it to the RA;
      (11) The RA forwards the digital identity to the WS;
      (12) Finally, the WS forwards the digital identity to the user.
      Although it is not explicitly depicted, we assume that several different
      SSL sessions are established between the User and the WS during the registration
      process. Nevertheless, this will be modelled in the subsequent formalizations.
      \label{fig:sequence}}
  \end{center}
\end{figure}

The messages sequence between the principals is depicted in Fig. \ref{fig:sequence},
where the continuous lines represent SSL protected communications (with server
authentication), and the dashed ones represent unprotected communications. The
unprotected communications are the messages 6.2 (email sent via SMTP) and the user 
accesses to the activation link through message 7.1. This last message is modelled
as unprotected because the communications are assumed not to be anonymized, and 
thus the attacker will be able to link the accessed URL and the user by listening 
on the communication channel. It is worth mentioning that in the diagram shown in 
Fig. \ref{fig:sequence} the activation email and the ticket are sent to the user as 
separate messages (messages 6.1 and 6.2 in the diagram). In addition, the user 
accesses the activation link and sends the ticket separately (messages 7.1 and 7.2
in the diagram). Nevertheless, in the formalization, both messages 6.1 and 6.2, 
and messages 7.1 and 7.2, are merged into messages 6 and 7, with unencrypted and
encrypted parts.

\begin{figure}[ht!]
  \begin{center}
    \includegraphics[width=0.80\textwidth]{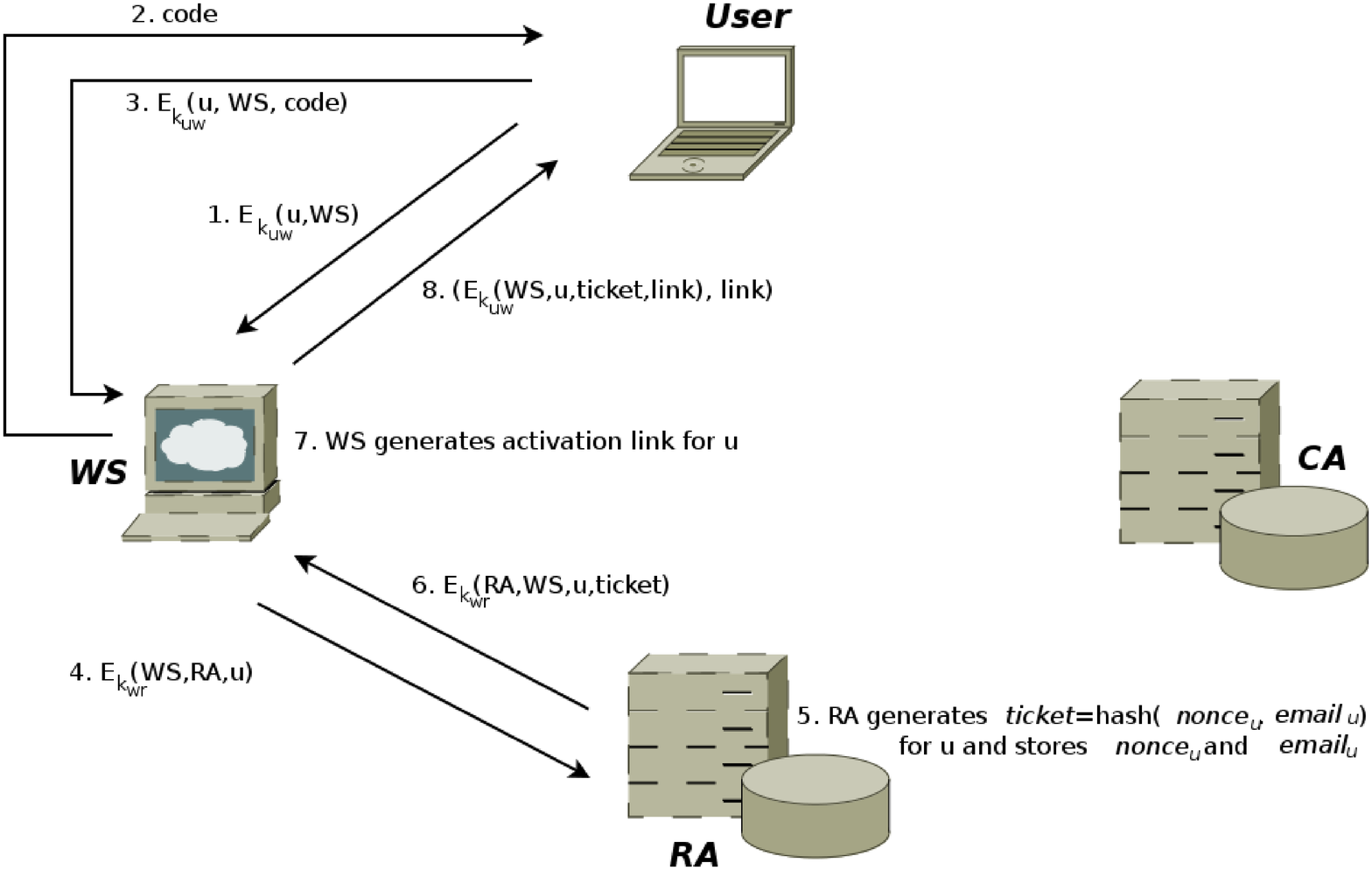}
    \caption{First half of the protocol: The user requests registration. 
      After validating the user data, the WS sends an SMS with a code 
      to the mobile phone number associated the new user. If the user
      correctly returns the code, the WS forwards the request to the
      RA, who generates a valid ticket linked to the user. When received
      the ticket, the WS also generates an activation link for that user,
      sending the ticket through SSL and the link via email. Here, $k_{uw}$
      represents the SSL key between the user and the WS, and $k_{wr}$ 
      represents the SSL key between the WS and the RA.\label{fig:informal1}}
  \end{center}
\end{figure}

\begin{figure}[ht!]
  \begin{center}
    \includegraphics[width=0.85\textwidth]{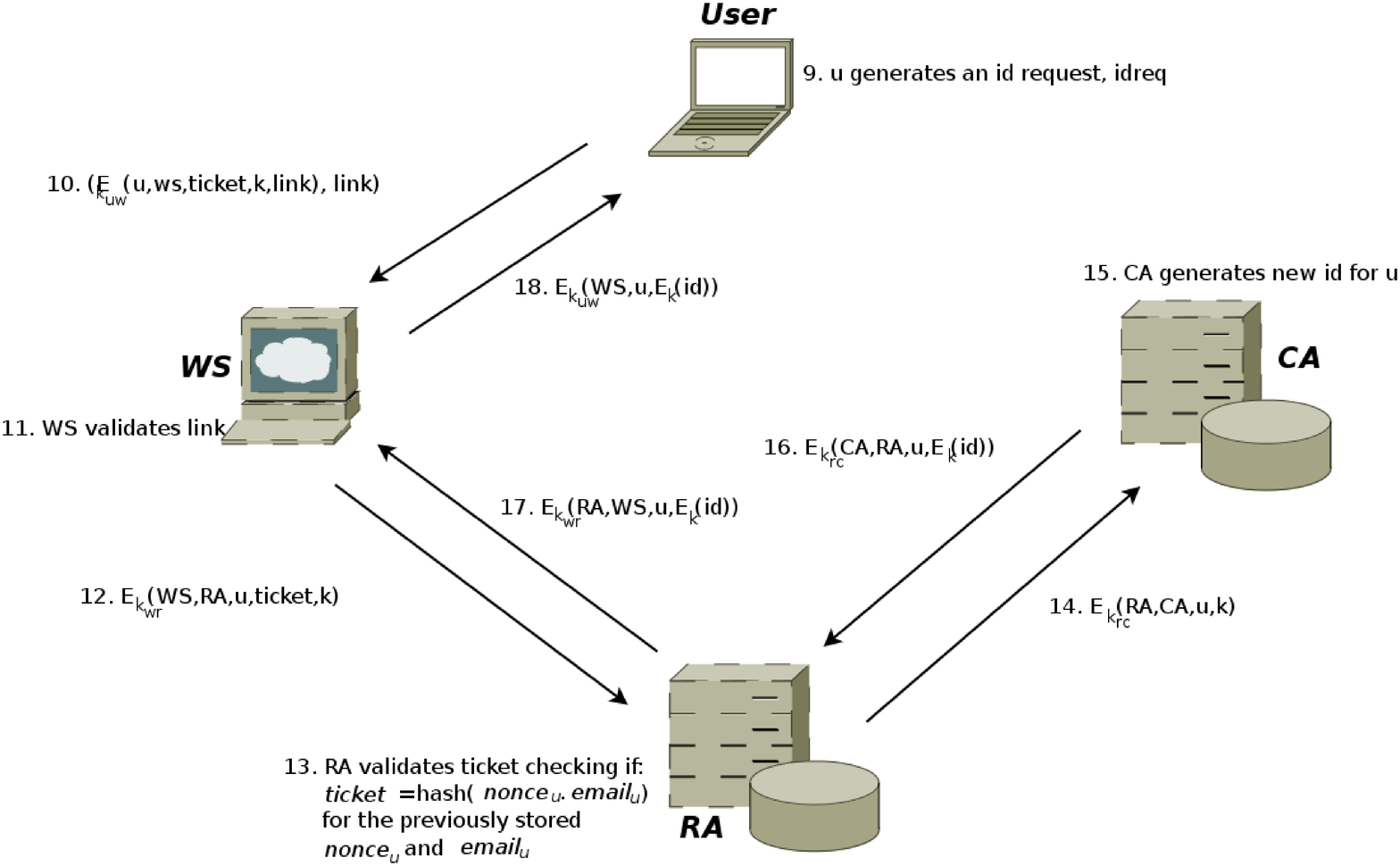}
    \caption{Second half of the protocol: The user generates a id request,
      accesses the activation link, and provides the registration ticket. After
      validating the user data, the WS forwards the ticket to the RA, who
      validates it using the nonce and email associated to that user, stored previously 
      (see step 3 of Fig. \ref{fig:informal1}), and requests the CA the generation 
      of a digital identity for that user. The CA generates the ID and sends it to 
      the user, passing through the RA and the WS. Here, $k_{uw}$ represents the SSL 
      key between the user and the WS, $k_{wr}$ represents the SSL key between the 
      WS and the RA, and $k_{rc}$ represents the SSL key between the RA and the CA.\label{fig:informal2}}
  \end{center}
\end{figure}

Concerning the principals distribution accross the network architecture,
obviously the user is an independent component by itself; but the 
network architecture of the three servers, i.e., WS, RA and CA is configurable.
This means that they can be all in the same physical server, each one at a 
different server, or WS and RA in one machine and the CA in another. This is
up to the system administrator. It is usually advised to keep the CA at a safe
place, even without direct connection to the internet, using a ``proxy'' between
it and the requesting users\footnote{See, for instance, the example given at
\url{http://www.ejbca.org/architecture.html}}. Typically, this proxy is the RA, which does have
access to the internet to receive registration requests. A sample network architecture with
each server at a different machine is depicted in Figs. \ref{fig:informal1}
and \ref{fig:informal2}, including the messages sent between them during the 
protocol. Both images correspond to the two main parts of the
protocol. In the first part, the user makes the registration request, obtaining
a ticket linked to his email and personal data along with an activation link; in the
second part, the user utilizes that ticket in order to obtain his final digital
identity.

\subsection{Robust multichannel authentication}
\label{ssec:multichannelauth}

In order to get registered into a typical interactive system, a user has to provide some personal 
data, which will
be used during the digital identity creation. This data typically consists at least on
a name and surname, email, and probably some other optional data (like postal 
address, phone numbers, or some national/organizational identification number). 
It is known that all this data is not too hard to obtain by combining internet
searches with techniques such as social engineering 
(\citep[Chapter 2]{anderson08}), or phishing \cite{oao11}. If the attacker
knows all the personal data of the person he wants to impersonate (and that
person is not already in the system), then the attacker just needs to start the
registration process and supply all the needed data.

\begin{figure}[ht!]
\begin{center}
\includegraphics[width=0.55\textwidth]{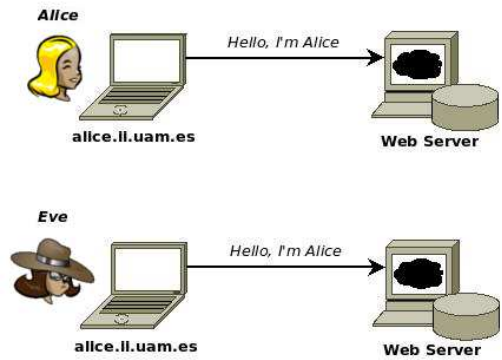}
\caption{Illustration of identity theft violating entity authentication: If 
\emph{Eve} gains control over Alice's host, the entity
authentication property is broken.\label{fig:identity-theft}}
\end{center}
\end{figure}

The problem exposed in the preceeding paragraph is exactly that of discerning
between \emph{authentication of origin} and \emph{entity authentication}, as
described in \citep[p.~8]{rs00}. The first refers to the fact of being sure \emph{where}
a message come from (for example, the machine \emph{alice.ii.uam.es}), while 
the second refers to \emph{who} sent it (for example \emph{Alice}). As settled
in \citep{rs00}, this difference is often unclear, maybe because the consequences
of one usually overlap with the consequences of the other. To be even clearer, let us
make use of a pedagogical example: Alice can send a message from her host 
\emph{alice.ii.uam.es}, which is legitimate. But if her host gets infected by 
\emph{Eve}, then the communicating entity will be \emph{Eve}, although the origin 
of the communication will still be \emph{alice.ii.uam.es}. This is a clear violation
of the entity authentication property (see Fig. \ref{fig:identity-theft}).

In \citep{ws05} the authors propose the use of multiple channels, each with different
security properties, in order to achieve authenticated communications. For example,
they make use of an extra channel with low transmission capacity but which cannot be tampered
with, although it may be subject to eavesdropping. This low capacity but secure
channel is used to transmit a single bit of information, telling if the verification
of the previous steps, carried on over a high capacity but unsecure channel,
has been successful or not. If the outcome of such verification
is positive, given their protocol, the communication will be \emph{origin authenticated}.
The authors in \citep{ar10,myt05} make a similar use of this multichannel combination
in order to achieve \emph{entity authentication}. Namely, they make use of mobile phones,
which are used to receive and/or send a \emph{One Time Pin} which will be returned
in order to confirm they are the legitimate owners of that phone number. This is basically
the same concept we used with emails, but with a subtle difference: users tend to control
more tightly their mobile phones than their emails.

This subtle fact is something worth considering in some depth. In \citep{ar10} it is 
reminded that there are basically three ways for verifying \emph{entity authentication}:
\emph{KBA} or Knowledge Based Authentication (something you know), \emph{TBA} or Token 
Based Authentication (something you have) and \emph{BBA} or Biometrics Based 
Authentication (something you are). Nevertheless, there is a key concept that is 
usually not even mentioned: the robustness of the authentication procedures is 
very dependant on the user perception about losing control of the supporting token
or device, since a possible lost could be expensive or annoying. This fact is 
obvious for BBA (everyone tends to avoid losing his own fingers or eyes), but it 
is no that obvious for KBA or TBA. Even more for KBA, because people using TBA 
are usually more security-concerned users. And precisely, KBA is what affects us, 
because the problem we have is that we assume someone's identity based on personal 
data required and, maybe, the fact of knowing the password to access an email account.

So, as considered above, we need to provide our KBA system with something that makes the
users think twice if their actions can lead to an annoying or undesirable situation,
while keeping the usability of the system. The systems proposed in \citep{ar10,myt05}
achieve precisely this property, providing a notorial improvement in security while
not (or almost not) reducing usability. The use of mobile phones for this purpose
suits our needs, as the fact of losing one's mobile phone leads to an
annoying situation, since mobile phones are currently perceived as a more critical personal
property than an email (maybe because they have physical presence). 
Therefore, the insertion of an additional step involving the sending
of a message containing a One Time Pin or something similar could help us in our
efforts to achieve \emph{entity authentication}. Moreover, as observed in \citep{ws05},
it will also help the user to avoid attacks to his account. If a user receives a message
at his mobile phone indicating that some action has been performed in her/his behalf, and
he is not responsible for it, then someone must be trying to impersonate him, and he
can just inform of the fact to thwart the attack. These reasons have led us to
incorporate the extra mobile phone channel in order to overcome all the previously
explained problems.

Nevertheless, there still exists a usability problem here. The perfect situation will be that in which the Service Provider
(in our case, the WS) knows in advance the mobile number of the user. If not, then
the attacker could simply provide the valid email of the user he wants to impersonate,
and then give his own mobile number. This may be an unfeasible task for many of the web sites, but, for example,
in the case of a university that wants its students and teachers to be registered
in an e-learning platform, this is not such a hard measure. The university will simply
need to ask their students/teachers to provide their mobile phone number when they get enrolled
or hired. And from then on, the users will be able to use their mobile phones to authenticate
themselves, and even to update their own mobile phone number. 

\subsection{Ticket generation}
\label{ssec:ticket}

As it can be seen in the sequence diagram depicted in Fig. \ref{fig:sequence} and
in Figs. \ref{fig:informal1} and \ref{fig:informal2}, the ticket generation
takes place at the RA after receiving its first message of the protocol. After
validating the user data, and checking that no user with that email has already
requested a ticket,\footnote{Other checks, like unicity of username, are done in the
WS, but the RA only has to process the tickets.} the RA creates a new one for the
requesting user. In order to do that, a nonce is generated. The nonce generation is of 
paramount importance to our protocol, as it is the element that allows us to
assert that a user successfully retrieving a digital identity is the one who
started the registration process. Therefore, we should adhere to the commitments of eSTREAM 
project\footnote{\url{http://www.ecrypt.eu.org/stream/}} to get an appropiate method 
to generate nonces. In our case, we have
used a chaos based Pseudo-Random Number Generator \citep{oggprm10}. Certainly, chaotic 
systems can be employed as skeleton of new, secure and efficient PRNGs (see \citep{amigo09}). 
Furthermore, cryptanalysis work in the field of chaos-based cryptography shows that 
security and efficiency can be achieved when there exists a proper combination of chaotic 
dynamics and the standards of conventional cryptography \citep{alvarez06a,arroyo08a,arroyo11,fp11}.
 
Once obtained the nonce, it is concatenated with the user email, and passed through a
one way hash function. This way, the user request has been univocally linked to the
freshly created ticket. This ticket is sent securely to the user through messages 3 and 4,
and it has to be returned later to the RA through messages 5 and 6 for the user to 
prove himself against the authority as the starter of the 
registration process. This very ticket is what prevents attacks like the one described
in the Introduction, as only the user who started the communication is the one
in posession of the correct ticket.

\subsection{Comparison with EBIA}
\label{ssec:comparison}

To be more clear about the reinforcements introduced by our protocol over EBIA, we
summarize them here, restating the attacks avoided with their incorporation.

First, with the incorporation of the authenticated mobile phone channel we ensure that
the users who succeed in the registration process are who they say they are. Note
that the mobile numbers are previously known by the registrar (the WS in our case).
In classic EBIA there is no way to avoid this, since there is no previous knowledge
about the users. Even if the registrar knew that a specific email is indeed associated
to a given user, since the email is inherently an insecure channel, it does not
guarantee the authenticity (nor the secrecy) of the messages sent during the protocol.

Second, let us suppose that we want to use classic EBIA, but with the extra mobile phone
authenticated channel. If, after the user succesfully provides the code sent to him 
by the mobile phone, we carry on using classic EBIA, then the MITM attack depicted in Section \ref{sec:securityreg} (Fig. \ref{fig:ebiaattack}) is still possible.
Therefore, the previous multichannel authentication by itself does not guarantee
anything if combined straightaway with classic EBIA. To reinforce this point, we have made use 
of tickets composed by nonces, which are used,
among other purposes, to avoid replay and MITM attacks and provide \emph{freshness} to
the protocol messages \citep{cgjpr08}. The concept of tickets has been widely
used in protocols and systems like Kerberos \citep{rfc4120}. In our protocol,
the user has to interact directly with the ticket created by the server.
In fact, once tickets are generated, the user stores it in her/his computer in order to 
prove her/his  when demanded. In our protocol, tickets are generated by 
applying a one-way hash function (in our case, SHA-2) to the concatenation of a nonce 
\citep[Chapter 3]{anderson08}, and the identifier of the email account of the requesting 
user. Nonces are generated from a chaos based \textbf{P}seudo \textbf{R}andom \textbf{N}umber \textbf{G}enerator (PRNG) (see 
\citep{oggprm10}), and the resulting ticket is sent through an SSL-protected channel when
the user applies for registration. Therefore, each ticket is univocally linked to a single request, 
since it is used as a receipt when asking for a digital identity and subsequently discarded. 
This way only who initiates the registration request will be able to complete it, 
solving the problem of sniffing the activation link.

\section{Security and usability analysis of the protocol}
\label{sec:analysis}

In the following sections we depict the secrecy and authenticity analysis by
using ProVerif. We start with the formalization of the protocol into
applied pi-calculus, which is used as input for ProVerif. After that, we will explain the
necessary conditions that our protocol needs to fulfill in order to convey
the required security properties. These conditions will be checked using ProVerif.
However, although secrecy and authenticity are main requirements of our protocol, but we must also assure 
usability. In Section \ref{ssec:usability} we consider some usability aspects, 
along with the results of a usability test carried out with real users.

\subsection{Formalization}
\label{ssec:formal}
Security protocols can be interpreted as concurrent systems, and thus they can be 
modelled using process calculi \citep{Cai:2010}. Pi-calculus
is one of those calculi, and it has been succesfully applied to model and analyse 
cryptographyc primitives. Nevertheless, the application of pi-calculus to this matter 
is not straightforward. It had to be modified, and thus the spi-calculus and applied 
pi-calculus were proposed. In this work we use Proverif, which is a 
practical implementation of applied pi-calculus. Next, we describe how our protocol
has been defined according to the notation of Proverif. Since 
the focus is on the application of the tool to our setup, the inner details of pi- 
and spi-calculus are not explained. The reader is referred to \citep{mpw92,ag97} for 
further details. 

We summarize the main components we used in Listing \ref{lst:pinotation}. In 
addition, in the formalization of the protocol the principals ($user,ws,ra,ca$)  
are noted using lower case letters to avoid confussion; when not 
referring to the formalization, principals are denoted with capital letters
(User, WS, RA, CA). Different messages are exchanged in each of those processes, 
which are labelled according to the pattern pattern $msgX$, with $X$ ranging from 
$1$ to $10$, plus an extra message $msgcode$ which is sent during the multichannel
authentication. In the four processes (User, WS, RA and CA) $net$ 
represents a public channel, which means that the attacker can eavesdrop on it, 
insert messages, and so on. The private channel $securemobilephonechannel$ 
is the one used for multichannel authentication and the also private channels
$privateSSLuserchannel$ and $privateSSLwschannel$ are used for SSL negotiation.

\begin{lstlisting}[frame=single,captionpos=b,
  caption={Basics of the pi-calculus notation.},
  label={lst:pinotation},
  escapechar=\%]
out(c,m) : Sends the message m through the channel c
in(c,m) : Receives the message m through the channel c
new n : Creates the %new% name n
%$P | Q$% : Given the processes %$P$% and %$Q$%, executes both %in% parallel 
%$!P$% : Given the %process% %$P$% replicates it any number of times
\end{lstlisting}

As mentioned above, our registration protocol is built upon four different processes,
simulating the principals of the protocol, plus two processes to simulate SSL 
negotiation between the User/attacker and the WS. We show below
these processes, plus the preamble with the functions, types and queries definitions. 
Additionally, the complete source code is available at \cite{code:chatsrp}.

\begin{description}
\item[Preamble (Listing \ref{lst:preamble})] The preamble is where the data types, 
  functions, events, queries, etc. are defined in ProVerif. In our case, we use 
  typical encryption/decryption functions, plus specialized functions for 
  encryption/decryption of digital identities. We also define private channels
  for SSL negotiation and SMS sending, and several selfdescriptive data types and
  free variables.
  \begin{lstlisting}[frame=single,captionpos=b,
    caption={ProVerif's preamble}.,
    label={lst:preamble},
    escapechar=\%, numbers=left]
    (** Data types **)
    type Host.
    type Nonce.
    type Key.
    type Tag.
    type Ticket.
    type Id.
    type Eid. (* Encrypted Id *)
    type Link.

    (** Channels **)

    (* A public channel. *)
    free net:channel. 
    
    (* A secure channel used for multichannel authentication *)
    free securemobilephonechannel:channel [private]. 

    (* A private channel for SSL negotiations. *)
    free privateSSLuserchannel:channel [private].

    (* A private channel for SSL negotiations. *)
    free privateSSLwschannel:channel [private]. 


    (** Principals **)
    free ws,ra,ca:Host. (* The servers names. *)

    (** Message tags. **)
    free msg1,msg2,msg3,msg4,msg5,msg6,msg7,msg8,msg9,msg10:Tag.
    free msgcode:Tag.

    (** Constructors and destructors **)

    (* Encryption and decryption *)
    fun encrypt(bitstring,Key):bitstring.
    reduc forall m:bitstring, k:Key; decrypt(encrypt(m,k),k) = m.

    (* ID encryption and decryption *)
    (* We use different functions for ID encryption/decryption
       for the secrecy queries *)
    fun encryptid(Id,Key):Eid. 
    reduc forall id:Id, k:Key; decryptid(encryptid(id,k),k) = id.

    (** Events **)
    event UserRequestsRegistration(Host).
    event WSSendsSMS(Host,Nonce).
    event UserProcessesSMS(Host,Nonce).
    event WSSendsLink(Host,Link).
    event RASendsTicket(Host,Ticket).
    event UserReceivesRegistrationData(Host,Ticket,Link).
    event UserReceivesId(Host,Id).
    event CASendsId(Host,Id).

    (** The queries. **)

    (* Secrecy *)

    (* The ID must remain secret both the encrypted and
       unencrypted versions*)
       query id:Id, k:Key; attacker(encryptid(new id,k)).%\label{lst:querysec1}%
       query id:Id; attacker(new id).%\label{lst:querysec2}%

    (* Authenticity *)

    (* Each time the CA generates and sends an ID to a 
       user h, it is because that user h has requested an 
       ID at least once *)
    query h:Host, id:Id; %\label{lst:queryauth1}%
    inj-event(CASendsId(h, id)) ==> 
              event(UserRequestsRegistration(h)).

    (* Each time a user receives an ID, it has been 
       sent by the CA *)
    query h:Host, id:Id; %\label{lst:queryauth2}%
    inj-event(UserReceivesId(h,id)) ==> 
              inj-event(CASendsId(h,id)).

    (* Each time a user processes an SMS, he has previously 
       requested it and it has been sent by the WS *)
    query h:Host, c:Nonce; %\label{lst:queryauth3}%
    inj-event(UserProcessesSMS(h,c)) ==> 
             (inj-event(WSSendsSMS(h,c)) %$\&\&$% 
              inj-event(UserRequestsRegistration(h))).

    (* Each time a user receives the registration data (link and 
       ticket), the link has been sent by the WS and the ticket 
       by the RA *)
    query h:Host, t:Ticket, l:Link; %\label{lst:queryauth4}%
    inj-event(UserReceivesRegistrationData(h,t,l)) ==> 
             (inj-event(WSSendsLink(h,l)) %$\&\&$%
              inj-event(RASendsTicket(h,t))).

    (* query h:Host, c:Nonce; 
       inj-event(WSSendsSMS(h,c)) ==>  %\label{lst:queryauthcomm}%
                 inj-event(UserRequestsRegistration(h)). *)
  \end{lstlisting}
\item[Process User (Listing \ref{lst:user})] The user first ``creates'' his 
  personal data. After proving his identity sending the code he receives by SMS, 
  he obtains the registration ticket and activation link, and uses them to finalize 
  the registration, receiving his digital identity. Note that three SSL key negotiations are
  performed. The first corresponds with the initial registration request, after which
  the user will receive an SMS. Subsequently, the user will respond with that code and
  receive the registration ticket and activation link. These messages that
  the user sends or receives correspond to the messages 1, 2, 3 and 8 of Fig.
  \ref{fig:informal1} and 10 and 18 of Fig. \ref{fig:informal2}.

\begin{lstlisting}[frame=single,captionpos=b,
  caption={Process User}.,
  label={lst:user},
  escapechar=\%, numbers=left]
(** The user process. **)
let userprocess =

  (* "Create" user data *)
  new u:Host;

  (* Launches the SSL key negotiation process for this 
     1st session *)
  out(privateSSLuserchannel,(u,ws));
  in(privateSSLuserchannel, (=u,=ws,kssluws1:Key));

  (* Tells WS to start the registration process *)
  event UserRequestsRegistration(u);
  out(net, encrypt((msg1,u,ws),kssluws1));

  (* After requesting to start the registration process, u  %\label{lst:usermcb}%
     receives a code via SMS *)
  in(securemobilephonechannel, (=u, code:Nonce));
  event UserProcessesSMS(u,code); 

  (* Send the code via the net channel to confirm identity *)
  out(privateSSLuserchannel,(u,ws));
  in(privateSSLuserchannel, (=u,=ws,kssluws2:Key));
  out(net, encrypt((msgcode, u, ws, code), kssluws2)); %\label{lst:usermce}%

  (* Receives the registration ticket and the activation link *)
  in(net,(cmsg4:bitstring,link:Link));
  let (=msg4, =ws, =u, ticket:Ticket, =link) = 
                              decrypt(cmsg4, kssluws2) in
  event UserReceivesRegistrationData(u,ticket,link);

  (* Launches the SSL key negotiation process for 
     this 3rd session *)
  out(privateSSLuserchannel,(u,ws));
  in(privateSSLuserchannel, (=u,=ws,kssluws3:Key));

  (* Accesses the activation link and provides the registration 
     ticket *)
  new k:Key;
  out(net, (encrypt((msg5,u,ws,ticket,link,k),kssluws3), link));

  (* Receives the digital identity *)
  in(net, cmsg10:bitstring);
  let (=msg10,=ws,=u,eid:Eid) = decrypt(cmsg10, kssluws3) in
  let id = decryptid(eid,k) in
  event UserReceivesId(u,id);
  0.
\end{lstlisting}
\item[Process WS (Listing \ref{lst:ws})] The WS receives the
    initial registration request from a user (or the attacker). Sends an
    SMS via the $securemobilephonechannel$ and after receiving back the
    code, it requests a registration ticket to the RA and generates an
    activation link. When the link is accessed, it makes a request for a 
    digital identity and finally forwards it to the user. The same SSL negotiations
    are perfomed here than in the user process. Note also that, since the
    WS and the RA are trusted third parties not in control of the attacker,
    we deliver them a symmetric key for securely communicating between them
    from the beginning. The messages sent or received by the WS correspond
    with the messages in steps 1, 2, 3, 4, 6 and 8 of Fig. \ref{fig:informal1}
    and 10, 12, 17 and 18 of Fig. \ref{fig:informal2}.
\begin{lstlisting}[frame=single,captionpos=b,
  caption={Process WS}.,
  label={lst:ws},
  escapechar=\%, numbers=left]
(** The ws process. **)
let wsprocess(kwsra:Key) =

  (* Waits until someone starts the 1st SSL key negotiation *)
  in(privateSSLwschannel, (u:Host,=ws,kssluws1:Key));

  (* Receives the request to start the registration process *)
  in(net, cmsg1:bitstring);
  let (=msg1,=u,=ws) = decrypt(cmsg1, kssluws1) in

  (* After establishing the SSL session with u, the ws, knows who  %\label{lst:wsmcb}%
     u is, and sends him a code via the extra secure channel 
     - that is, by SMS *)
  new code:Nonce;
  event WSSendsSMS(u,code);
  out(securemobilephonechannel, (u, code));

  (* Receives the keys of the 2nd SSL key negotiation *)
  in(privateSSLuserchannel, (=u,=ws,kssluws2:Key));
  in(net, cmsgcode:bitstring);
  let (=msgcode, =u, =ws, =code) = decrypt(cmsgcode,kssluws2) in %\label{lst:wsmce}%

  (* Request a registration ticket to the RA, creates an 
     activation link and forward them to the user *)
  out(net, encrypt((msg2,ws,ra,u),kwsra));
  in(net, cmsg3:bitstring);
  let (=msg3,=ra,=ws,=u,ticket:Ticket) = decrypt(cmsg3,kwsra) in
  new link:Link;
  event WSSendsLink(u,link);
  out(net, (encrypt((msg4,ws,u,ticket,link),kssluws2), link));

  (* Waits until the 3rd SSL key negotiation *)
  in(privateSSLwschannel, (=u,=ws,kssluws3:Key));

  (* Processes the access to the activation link and forwards 
     the received registration ticket to the RA *)
  in(net, (cmsg5:bitstring,=link));
  let (=msg5,=u,=ws,ticket':Ticket,=link,k:Key) = 
                                 decrypt(cmsg5, kssluws3) in
  out(net, encrypt((msg6,ws,ra,u,ticket',k),kwsra));

  (* Receives the digital identity and forwards it to the user *)
  in(net, cmsg9:bitstring);
  let (=msg9,=ra,=ws,=u,eid:Eid) = decrypt(cmsg9, kwsra) in
  out(net, encrypt((msg10,ws,u,eid),kssluws3)).
\end{lstlisting}
\item[Process RA (Listing \ref{lst:ra})] The RA receives a request from a user via the WS.
    After validating the user data, the RA creates a ticket for that user and sends it to the WS.
    When the ticket is received again, compares it to the one previously created. If they
    match, the RA makes a request to the CA for the generation of a digital identity for that user.
    In the end, the RA forwards the digital identity to the WS. The RA shares secret
    keys with the WS and the CA, pre-established from the beginning. The messages involving
    the RA correspond with messages 4 and 6 of Fig. \ref{fig:informal1} and 12, 14, 16 and 17
    of Fig. \ref{fig:informal2}.
    
\begin{lstlisting}[frame=single,captionpos=b,
  caption={Process RA}.,
  label={lst:ra},
  escapechar=\%, numbers=left]
(** The ra process. **)
let raprocess(kwsra:Key, kraca:Key) =

  (* Receives the request for a new registration ticket *)
  in(net, cmsg2:bitstring);
  let (=msg2,=ws,=ra,u:Host) = decrypt(cmsg2, kwsra) in

  (* Create the ticket and send it to the WS *)
  new ticket:Ticket;
  event RASendsTicket(u,ticket);
  out(net, encrypt((msg3,ra,ws,u,ticket),kwsra));

  (* Receive a supposedly legitimate registration ticket *)
  in(net, cmsg6:bitstring);
  let (=msg6,=ws,=ra,=u,ticket':Ticket,k:Key) = 
                                    decrypt(cmsg6,kwsra) in

  (* Checks the ticket. If everything is OK, request the CA 
     a new digital identity for the corresponding user *)
  if ticket' = ticket then
  out(net, encrypt((msg7,ra,ca,u,k),kraca));

  (* Receive and forward the digital identity *)
  in(net,cmsg8:bitstring);
  let (=msg8,=ca,=ra,=u,eid:Eid) = decrypt(cmsg8, kraca) in
  out(net, encrypt((msg9,ra,ws,u,eid),kwsra)).
\end{lstlisting}
\item[Process CA (Listing \ref{lst:ca})] The CA only receives requests from the RA.
    For each request, the CA generates a digital identity with the user data received and
    sends the digital identity to the RA. The CA therefore shares a secret key with the RA,
    also pre-established from the beginning. These messages correspond with the messages 16
    and 17 of Fig. \ref{fig:informal2}.
\begin{lstlisting}[frame=single,captionpos=b,
  caption={Process CA}.,
  label={lst:ca},
  escapechar=\%, numbers=left]
(** The ca process. **)
let caprocess(kraca:Key) = 

    (* Receives a request of a new digital identity *)
    in(net,cmsg7:bitstring);
    let (=msg7,=ra,=ca,u:Host,k:Key) = decrypt(cmsg7, kraca) in

    (* Create the digital identity and send it to the RA *)
    new id:Id;
    let eid = encryptid(id,k) in
    event CASendsId(u,id);
    out(net, encrypt((msg8,ca,ra,u,eid),kraca)).
\end{lstlisting}
\item[SSL negotiation processes (Listing \ref{lst:sslnegotiation})] The user communicates 
  with the WS through SSL channels. The corresponding keys are established using
  the process $sslkeynegotiationprocess$. We simulate the capability of the attacker
  to establish SSL sessions with the WS using the process $sslbypass$, which creates
  a key, makes it public, and sends it to the WS like a key created with a legitimate
  user.  
\begin{lstlisting}[frame=single,captionpos=b,
  caption={Processes for SSL negotiation}.,
  label={lst:sslnegotiation},
  escapechar=\%, numbers=left]
(** SSL negotiation process **)
 (* - used just between the User and WS *)
let sslkeynegotiationprocess =

    (* u is the user who is supposedly establishing a SSL 
       session with WS a is who really is establishing the 
       SSL session WS *)
    in(privateSSLuserchannel, (u:Host,=ws));

    (* Creates the new SSL session key *)
    new kssluws:Key;

    (* Send the new SSL key through the privatesslchannel 
       channel, to let WS know the new key - and the user, in
       case it is not the attacker who establishes the session *)
    out(privateSSLuserchannel, (u,ws,kssluws));
    out(privateSSLwschannel, (u,ws,kssluws)).

(** This process allows the attacker to establish SSL sessions 
    with the ws **)
let sslbypass =

    in(net, (h:Host,=ws));
    new ksslhws:Key;
    out(net, ksslhws);	
    out(privateSSLwschannel, (h,ws,ksslhws)).
\end{lstlisting}
\item[Main process (Listing \ref{lst:main})] The previous principals are called from the
  main process. In this process the secret shared keys between the trusted third
  parties are created. Obviously, we allow several replications for each principal.
\begin{lstlisting}[frame=single,captionpos=b,
  caption={Main process}.,
  label={lst:main},
  escapechar=\%, numbers=left]
(** The system. **) 
process
  
  (* Since the WS, RA and CA are trusted third parties, we can
     establish symmetric keys for communications between them *)

  new kwsra:Key; 
  new kraca:Key;

  (* Launch all the processes *)
  ( (!userprocess)                  (* Users *)
    | (!sslkeynegotiationprocess)   (* SSL negotiation *)
    | (!sslbypass)                  (* Attacker SSL bypass *)
    | (!wsprocess(kwsra))           (* Moodle Server *)
    | (!raprocess(kwsra,kraca))     (* RA *)
    | (!caprocess(kraca))	    (* CA *)
  )
\end{lstlisting}
\end{description}

These four processes make up the fourth step in our methodology and, consequently, 
we can proceed with the security analysis of the registration protocol. In the 
next section we prove the security properties required in Section \ref{ssec:goals}.

\subsection{A quick introduction to ProVerif}
\label{ssec:proverif}

ProVerif\footnote{\url{http://www.proverif.ens.fr/}} is an
automatic formal verifier of security properties for communication protocols 
\citep{blanchet01}. It accepts a formal definition of a protocol as pi-calculus 
instructions or Horn clauses \citep{horn51}, and a set of secrecy and/or events' 
correspondence queries to be proved. In both cases (pi-calculus and Horn clauses), 
the input is transformed into a set of Horn clauses which is completed and refined
in a series of iterations. After the completion, a goal-directed depth-first search 
is carried out to prove the specified queries. Roughly speaking, secrecy verification 
is performed by applying ProVerif to assess if any of the implied terms can be 
derived from the obtained ruleset. To prove authenticity, we use Proverif to test 
the possibility of creating an execution trace of the protocol in which any of the 
events correspondences is broken. That is: if event A is supposed to happen before 
event B (because it is said so in a correspondence query) and ProVerif finds a 
trace in which that does not happen, then the correspondence is broken. Note that, 
by definition, an attacker is a process in which no event occurs \cite{blanchet02}.

To ask ProVerif to prove if a protocol keeps a given property, it has to be 
explicitly inquired about it using proper instructions. The output of Proverif 
indicates if the queried security properties do or do not hold. If ProVerif founds
that the property related to a specific query is false, it will show, along with an
attack trace, something like\footnote{We exclude some extra identifiers appended by 
ProVerif to the names, which are used to differentiate between different runs of the 
processes. Take into account that a single process can be executed several times, and ProVerif has to 
distinguish the different variables created in each run.}:

\begin{quote}
\begin{verbatim}
The attacker has the message property.
A trace has been found.
RESULT not attacker:property is false.
\end{verbatim}
\end{quote}

\subsection{Secrecy}
\label{ssec:secrecy}

In ProVerif, secrecy properties can be checked with queries of the form:

\begin{quote}
\begin{verbatim}
query attacker(T).
\end{verbatim}
\end{quote}

Which is the way to ask ProVerif to test if the attacker can gain knowledge
of some term \texttt{T}.

Since the aim of our protocol is to convey private digital identities to new 
users, these identities must remain secret. Moreover, in our protocol, the 
digital identities are sent encrypted under a symmetric key specified by the 
user. Therefore, even if this encrypted versions of the digital identities 
fall into the hands of an attacker, the secrecy is considered broken. More 
formally:

\begin{definition}
  CHAT-SRP preserves secrecy if neither the created digital identities nor their
  corresponding encrypted versions are disclosed to attackers.
  \label{def:sec}
\end{definition}

We check this in ProVerif with the queries in lines \ref{lst:querysec1} and 
\ref{lst:querysec2} of Listing \ref{lst:preamble}. As a result, ProVerif informs 
that the corresponding secrecy properties are held. Therefore, given ProVerif's 
soundness property, we have the guarantee that no attacker gains knowledge of the 
digital identities nor their encyrpted versions.

\subsection{Authenticity}
\label{ssec:authenticity}

The way of proving authenticity properties with ProVerif is by means of correspondence
assertions \cite{wl93,blanchet10}. These correspondence assertions allow us to check, for instance,
if a given event $e$ always preceeds another event $e'$. Since the attackers are, by
definition, processes without events, if an event occurs, it must have been invoked
by a \emph{legitimate} process. Moreover, we can include variables in these events, 
to see if a given variable has the same value in an event $e$ than in another event 
$e'$.

Since, after a successful execution of our protocol, a new user acquires a digital
identity which will be linked to him in the interactive platform, this identity 
must fulfill authenticity requirements (besides being kept secret). Specifically, 
we require that if the CA sends a digital identity to an user, then the legitimate 
corresponding user has requested it. This guarantees that no user receives a
digital identity of a different user, and that only legitimate users can request
a digital identity. We also need to guarantee that, every time a user receives
a digital identity, it has been created by the CA. These requirements are more 
formally stated in definitions \ref{def:auth1} and \ref{def:auth2}, respectively,
corresponding to the lines \ref{lst:queryauth1} and \ref{lst:queryauth2} in Listing
\ref{lst:preamble}:

\begin{definition}
  If the CA issues a digital identity intended for a user $h$, then the legitimate 
  user $h$ must have requested it, at least once.
  \label{def:auth1}
\end{definition}

\begin{definition}
  If a user receives a digital identity $id$, that digital identity $id$ must
  have been issued by the Certification Authority.
  \label{def:auth2}
\end{definition}

While the second requirement is held even if we do not make use of multichannel
authentication, the first requirement does need this additional mechanism. To
see this, one can delete lines \ref{lst:usermcb} to \ref{lst:usermce} in Listing
\ref{lst:user} and lines \ref{lst:wsmcb} to \ref{lst:wsmce} in Listing \ref{lst:ws}
in order to eliminate the multichannel authentication in the formalization. After
doing so, ProVerif founds a trace that contradicts the property required in 
definition \ref{def:auth1}. In short, after establishing an SSL session with the
WS, the attacker provides all the required data in order to successfully impersonate
some user. As a result, the attacker finally obtains an illegitimate, but valid,
digital identity. This was the attack described in the first paragraph of section 
\ref{ssec:multichannelauth}.

Therefore, this helps us see why we need to share some information with verified 
authenticity. The fact of using multichannel authentication enforces
the robustness of our protocol. In any case, this multichannel exchange also needs
to be verified. Namely, we will require that each time a user processes a code
received via SMS, that same user has previously requested registration and the 
WS has send to him the same code. A logical consequence of this, captured in
another requirement is that, each time a user receives a registration ticket
and an activation link, then the RA has generated that same registration ticket
and the WS has created the same activation link. These two requirements are
formally stated in definitions \ref{def:auth3} and \ref{def:auth4}, respectively,
and are coded in lines \ref{lst:queryauth3} and \ref{lst:queryauth4} in Listing
\ref{lst:preamble}:

\begin{definition}
  Each time a user $h$ processes an SMS with a code $c$, then the WS has send to
  that user $h$ an SMS containing the code $c$, and the user $h$ has also requested to be
  registered in the system.
  \label{def:auth3}
\end{definition}

\begin{definition}
  Each time a user $h$ receives a registration ticket $t$ and an activation link
  $l$, then previously the RA has created the ticket $t$, the WS has created the
  link $l$, and both have sent them to the same user $h$.
  \label{def:auth4}
\end{definition}

After running ProVerif with the previous correspondence assertions, it informs
that the associated properties are kept. Again, ProVerif soundness guarantees
the result.

Nonetheless, one more point is worth to be noted. In the multichannel alternative, the
attacker is yet capable to start a registration process by providing the required
data to the WS. But, in this case the WS will check if it knows the mobile phone
number of the user who is allegedly requesting registration. If possitive, the WS
will send an SMS to the known number. When the real user receives the SMS,
he will know that someone is trying to impersonate him, and he will ignore the
SMS, and even inform the corresponding authority. Even though it was the attacker who 
started the registration, and not the real user, the WS will be sending an SMS to the 
correct number, and the registration will not succeed. This fact can be seen if we uncomment
the query in line \ref{lst:queryauthcomm} of Listing \ref{lst:preamble}, which
informs us that a given user will not always process an SMS with a given code, 
even though the WS has previously sent it. This is not a weakness of the protocol.
In fact, it is rather telling us that the multichannel authentication is working,
because it serves to avoid impersonation.

\subsection{Usability}
\label{ssec:usability}

As we have already said, usability and security inevitably lead to a tradeoff where
the system designers. must find an adequate equilibrium. In \citep{gg05},
the authors pinpoint some real situations where a system or protocol was not widely used, 
although a high level of security was implemented. All the reasons there stated concern 
usability: interface complexity, difficult configuration, etc. Moreover, 
they give examples of secure (but not usable) systems replaced by usable (but less 
secure) ones, even though the original purpose of both systems slightly differed.

It is even astonishing that, among the most known cryptographic principles, those written 
by Kerchkoffs in 1883 there is one principle referring this matter (see \citep[page 12]{kerckhoffs1883}). 
It is the last of 6 principles, and states:

\begin{principle}[Kerckhoffs]
Lastly, it is needed, given the circumstances that command its application, for the system to 
be easily usable, not requiring mental strain nor the knowledge of a long series of rules.
\end{principle}

The problem here is that
there is no exact or even approximate equation, theory or whatever that could tell if
a system is $100\%$ usable. Therefore, the approach we have taken here is to base our
work in concepts that are familiar to the vast majority of the potential users of our
protocol (that will ease its understanding). Regarding the base and familiar concepts
of our system, we can resume the protocol in three of them:

\begin{itemize}
\item[1.] \textbf{EBIA}: Email Based Identification and Authentication (\citep{garfinkel03}). 
It is the starting point we took for our protocol. It links each user identity to his or her email
account. As emails are something to which everyone is used nowadays, it keeps the
property of familiarity.
\item[2.] \textbf{Tickets}: They are used to strengthen the basic EBIA system. 
Created from a chaotically-generated random number concatenated with some user dependent data.
They serve to univocally link each specific user with a single registration request. They are used
as a nonces, so once used, they are deleted in order to avoid multiple registrations.
As we said in \citep{dar11}, everyone is used to get tickets in real life and keep
them as a recepit to prove something later (e.g., to return clothes). Therefore,
it also keeps the property of familiarity.
\item[3.] \textbf{SMS}: When used the multichannel authentication combining emails
and SMS, the user will receive a short PIN into his mobile phone.
Although this might not be a very common practice yet, it is gaining popularity,
and mobile phones take a central roles in technology users nowadays. It seems then
that, when used, this option still keeps the property of familiarity.
\end{itemize}

Keeping the familiarity, what remains is
to create a friendly, self-descriptive, intuitive and as simple as possible interface,
and that will not depend on the protocol itself.  Nevertheless, the best, and maybe the only,
way to foresee the users acceptance, is to test the system with real and potential
users. Therefore, as a proof of concept, we performed some trials with real and potential 
users (students at our university). In these trials, the users were given a global description
of the system. After the introduction, they had to register themselves in the Moodle
test platform\footnote{They also had to digitally sign an online exam, but we do not treat
that matter here.}. The registration process tested did not include the two-factor authentication
scheme including SMS, but just the activation email and the ticket. At last, they were 
requested to fill up a questionnaire with the questions related to the adjectives in Table 
\ref{tab:trials}. From those tests we gained very valuable feedback and quite positive 
opinions, which are summarized in Table \ref{tab:trials}.

\begin{table}[ht!]
\centering
\begin{tabular}{|c|c|}
%\hline
%\multicolumn{2}{|c|}{\cellcolor{black}\textcolor{white}{Table \ref{tab:trials}: Trials results}}\\
\hline
Adjective&Mean score (1 to 5)\\
\hline
Simple&3.60\\
Quick&4.20\\
Intuitive&3.19\\
Well developed&4.26\\
Secure&4.86\\
Useful&4.65\\
Trustworthy&4.73\\
Advisable for use&4.73\\
\hline
\end{tabular}
\caption{Results of the trials performed with real users to measure the usability of the protocol.
For each question, concerning an adjective describing the system, the user had to answer with a
numerical score, ranging from 1 to 5, where 1 meant ``Completely disagree'' and 5 meant
``Completely agree''. A total of 15 persons took the test.\label{tab:trials}}
\end{table} 

In the results, the first four adjectives can be considered as directly related to usability. 
The last four, although not directly related, will get bad scores may the user not understand
or not know how to use the system, so they can also be seen as indirect measures of usability.
Moreover, the users still have a sense of being using a secure system, which increases their
confidence on the system, considering it trustworthy and advisable for use, may they find 
themselves in a situation in which they had to decide whether to incorporate the protocol or
not. This latter property is quite important, because it highlights that the users admit a
slight loss in usability (that is inevitable) in order to gain in security. Nevertheless, we
have obtained very valuable feedback during the tests, regarding usability, and hope to
be able to improve it.

\section{Conclusion}
\label{sec:conclusion}

% Work summary and main contributions
In this work we have presented a new registration protocol for interactive and 
collaborative platforms, CHAT-SRP (\textbf{CHA}os based \textbf{T}ickets-\textbf{S}ecure 
\textbf{R}egistration \textbf{P}rotocol), which provides a very reasonable tradeoff
between security and usability.

We have taken as starting point the EBIA model, which provides us a 
good reference from the usability perspective, since it is the most widely
used protocol for registration in interactive platforms. As for its security, 
we have circumvented the two main problems inherent to these protocols. Namely, 
we prevent impersonation (with an authenticated extra channel) and MITM attacks 
(with the incorporation of registration tickets). Nevertheless, we still required to
formally verify that our measures successfully avoid those weaknesses. Therefore,
to evaluate the security of the protocol we have followed a 
methodology divided in interconnected phases. Obviously, the first
one was to determine the protocol goals, which for us is to provide the new users
with a digital identity. Once known that, we have established a general security 
model (the Dolev-Yao model, which assumes that the protocol uses perfect and unbreakable 
cryptographic primitives), in order to be available later to check if the required 
properties are held. After setting the general security model, we have defined the
security requirements. In our case,
we required secrecy and authenticity for the distributed digital identity. The next
logical step is to formalize the protocol in a language suitable for being analyzed
with formal tools, task undertaken in Section \ref{sec:protocol}. We have used the 
applied pi-calculus for that purpose. At last, it remains to formally verify that the required 
properties are held. We have made use of ProVerif to verify them. In Section 
\ref{sec:analysis} we proved these properties along with a usability analysis of the 
protocol.

As we have proved using ProVerif (Section \ref{ssec:secrecy}), our protocol keeps 
the secrecy of the digital identities, and it also keeps the authenticity property
(Section \ref{ssec:authenticity}) when facing both internal and external attackers. 
Of course, both according to the security model we have assumed. It does so by the
combination of a multichannel authentication method (with some previous knowledge
in the form of a mobile phone number) and of a registration ticket we have introduced 
to univocally link each user with his corresponding request during the whole process. 
This ticket is created by introducing 
chaos based pseudo-random nonces, concatenated with the user email, which is used
as preliminary identifier. 

We have made use of concepts familiar to the users: EBIA, tickets, and SMS. Nevertheless, for
measuring the usability there is no more precise science than usability tests 
involving real users. For that purpose, we have carried out some trials with potential
users. For the tests, we incorporated our protocol (without the two-factor SMS
authentication) into a Moodle platform, which is the perfect scenario for testing our
protocol. As a result of the distribution of digital identities, the users
who took part in the test were able to deliver digitally signed online exams, and
became the first users to do so in our university. After the tests, they filled up
a questionnaire asking about their experience with the system. The obtained results 
have been shown in Section \ref{ssec:usability}. From them we can conclude that the 
acceptance is high, although we can still improve it. Besides, we received very valuable
feedback.

% Discussion
As a result of providing digital identities to users of such collaborative and 
interactive systems, a full range of new cryptographically robust funcionality is
available for them. This provides greater security than using just SSL protected
communications, since the users can now authenticate themselves with their digital
identities. They could do it with SSL, may they have a digital identity, but they
would not have one until our protocol (or other similar) provides them with one.
Also, the fact of distributing digital identities in a standard format (like X.509),
opens very interesting posibilities, like seamlessly adding advanced
functionalities like privacy-friendly authentication and anonymity \cite{kwon11,bcly08}. 
Moreover, as we have seen from the tests, the users 
do perceive a greater security when they have their personal digital identity 
available (justified, as we have seen). This makes them have a greater trust in 
the system, increasing their acceptance, as they usually have to provide 
sensitive data to this kind of systems.

\section*{Acknowledgments}
This work was supported by the UAM project of Teaching Innovation and the Spanish Government
project TIN2010-19607. The work of David Arroyo was supported by a Juan de la 
Cierva fellowship from the Ministerio de Ciencia e Innovaci\'on of Spain.

\bibliographystyle{plain}
\bibliography{moodle_secure_verif}

\end{document}